# A spectral phase modulation transfer function for dispersive four-wave mixing

LINSHAN SUN,[1,*] AND SERGIO CARBAJO[1]

[1]*Department of Electrical & Computer Engineering, University of California Los Angeles, Los Angeles, CA 90095,*
**lssun@ucla.edu*

**Abstract:** Indirect control of ultraviolet (UV) pulse phase through nonlinear frequency conversion is attractive when direct UV pulse shaping is limited by material loss, dispersion, and damage threshold. Here we cast dispersive four-wave mixing (DFWM) as a pump-conditioned spectral kernel and show that, in a locally one-to-one mapping regime, the signal-to-idler conversion admits a practical transfer function description. Starting from the exact frequency domain expression, we rewrite the idler field as a linear operator acting on the conjugated signal spectrum, with a two-frequency kernel set by the pump self-convolution and phase matching. Linearization around a reference operating point then yields a spectral phase-response kernel for small input perturbations. By probing this response with sinusoidal spectral-phase modulation of different spatial frequencies, we define a spectral phase-modulation transfer function (SPMTF), an MTF-like measure of the phase-transfer bandwidth of the nonlinear interaction. Simulations with different pump group-delay dispersion (GDD) values produce distinct SPMTF curves, showing that pump chirp directly controls how much fine spectral-phase structure survives the conversion. This framework provides a simple way to compare operating conditions and identify regimes favorable for programmable NIR-to-UV phase transfer.

1. **Introduction**

Programmable control of ultraviolet waveforms remains difficult, even though many ultrafast experiments would benefit from it[1–3]. In practice, direct UV pulse shaping is constrained by the limited availability and performance of UV-compatible components, as well as by stronger loss, dispersion, and lower damage thresholds than in the visible or near-infrared. One way around this problem is to shape a near-infrared (NIR) pulse, where pulse-shaping technology is mature, and transfer that structure to the UV through nonlinear frequency conversion. Recent progress in short-pulse UV/DUV generation and manipulation has made this indirect route increasingly realistic[4–7].

What is still missing is a simple metric for comparing how well different DFWM conditions transfer spectral phase. Most demonstrations are still discussed case by case, which makes it hard to compare operating points or to state where the useful phase-transfer bandwidth begins to fail. Here, we introduce a spectral phase-modulation transfer function (SPMTF) for that purpose. Starting from the exact frequency-domain expression for DFWM, we rewrite the signal-to-idler mapping as a pump-conditioned transfer kernel, linearize it around a reference spectrum, and probe the response with sinusoidal spectral-phase perturbations of varying modulation frequency. The resulting SPMTF compresses the phase-transfer behavior into a single curve and makes the role of pump chirp easy to visualize[6,7].

2. **Theory**

**Kernel formulation of DFWM**

We begin from the frequency-domain expression for degenerate four-wave mixing, in which the generated idler field is written as a double integral over the two pump frequencies. Under the undepleted-pump approximation and an instantaneous third-order nonlinear response, the idler spectral field can be written as

$$\widetilde{E}_i(\Omega) = C \iint \widetilde{E}_p(\omega_1)\widetilde{E}_p(\omega_2)\widetilde{E}_s^*(\omega_1 + \omega_2 - \Omega)\,\eta(\omega_1, \omega_2, \omega_1 + \omega_2 - \Omega)\,d\omega_1\,d\omega_2, \qquad (1)$$

where $\tilde{E}_p$, $\tilde{E}_s$, and $\tilde{E}_i$ are the pump, signal, and idler spectral fields, respectively, $C$ is a constant prefactor, and $\eta$ represents the phase-matching and propagation weighting. Equation (1) makes explicit that the DFWM idler is intrinsically a three-field spectral mixing process rather than a conventional linear filter.

To obtain a transfer-like description, we rewrite the signal term as

$$\tilde{E}_s^*(\omega_1 + \omega_2 - \Omega) = \int \tilde{E}_s^* \delta[\omega - (\omega_1 + \omega_2 - \Omega)]d\omega. \tag{2}$$

Substituting Eq. (2) into Eq. (1) and changing the order of integration yields

$$\tilde{E}_i(\Omega) = \int H_E(\Omega, \omega) \tilde{E}_s^*(\Omega) d\omega, \tag{3}$$

with the two-frequency kernel

$$H_E(\Omega, \omega) = C \iint \tilde{E}_p(\omega_1) \tilde{E}_p(\omega_2) \eta(\omega_1, \omega_2, \omega) \delta(\Omega + \omega - \omega_1 - \omega_2)] d\omega_1 d\omega_2, \tag{4}$$

Using the delta function to eliminate one integration variable gives

$$H_E(\Omega, \omega) = C \int \tilde{E}_p(\omega_1) \tilde{E}_p(\Omega + \omega - \omega_1) \eta(\omega_1, \Omega + \omega - \omega_1, \omega) d\omega_1, \tag{5}$$

Equation (5) shows that, once the pump is treated as fixed, the signal-to-idler mapping can be viewed as a linear operation acting on the conjugated signal field. The kernel is determined by the pump self-convolution together with the phase-matching weight. In this sense, the original double integral is not removed; it is absorbed into a two-frequency transfer kernel.

In the idealized regime considered here, we ignore self-phase modulation, group-velocity mismatch, plasma generation, and other propagation-induced distortions, and we assume that the phase-matching factor varies slowly across the spectral window of interest. For a strongly chirped pump, the interaction can then approach a local one-to-one time-frequency mapping, so that the two-dimensional kernel becomes concentrated around a signal-to-idler trajectory. The field transfer is then well approximated by

$$H_E(\Omega, \omega) \approx G_E(\Omega; \phi_{2,p}) \delta(\omega - \omega_m(\Omega)), \tag{6}$$

so that

$$\tilde{E}_i(\Omega) \approx G_E(\Omega; \phi_{2,p}) \tilde{E}_s^*(\omega_m(\Omega)). \tag{7}$$

Here $G_E(\Omega; \phi_{2,p})$ is an effective one-dimensional spectral transfer function and $\omega_m(\Omega)$ is the local mapping between signal and idler frequencies. Near the central frequencies in degenerate DFWM, this mapping is approximately antisymmetric, consistent with $\omega_i \approx 2\omega_p - \omega_s$. The pump GDD enters through its control of the local time-frequency selectivity of the interaction.

**Spectral phase-transfer kernel and SPMTF**

To quantify phase transfer rather than only field transfer, we consider a small spectral-phase perturbation applied to a reference signal spectrum. Let the signal field be written as

$$\tilde{E}_s(\omega) = \tilde{E}_{s0}(\omega)exp\{i\delta\phi_s(\omega)\} \approx \tilde{E}_{s0}(\omega)[1 + i\delta\phi_s(\omega)] \tag{8}$$

where $\tilde{E}_{s0}(\omega)$ is the reference field, and $\delta\phi_s(\omega)$ is a small phase perturbation. Its complex conjugate is then

$$\tilde{E}_s^*(\omega) = \tilde{E}_{s0}^*(\omega)[1 - i\delta\phi_s(\omega)] \tag{9}$$

Substituting Eq. (9) into Eq. (3), the output idler field can be written as

$$\tilde{E}_i(\Omega) = \tilde{E}_{i0}(\Omega) + \delta\tilde{E}_i(\Omega) \tag{10}$$

with the unperturbed output

$$\tilde{E}_{i0}(\Omega) = \int H_E(\Omega, \omega)\tilde{E}_{s0}^*(\Omega)d\omega, \tag{11}$$

and the first-order perturbation

$$\delta\tilde{E}_i(\Omega) = -i\int H_E(\Omega, \omega)\tilde{E}_{s0}^*(\Omega)\delta\phi_s(\omega)d\omega, \tag{12}$$

For sufficiently small perturbations, the induced idler phase change is

$$\delta\phi_i(\omega) \approx \Im\left[\frac{\delta\tilde{E}_i(\Omega)}{\delta(\Omega)}\right] \tag{13}$$

This yields a linear phase-response relation of the form

$$\delta\tilde{E}_i(\Omega) = \int T_\phi(\Omega, \omega)\delta\phi_s(\omega)d\omega, \tag{14}$$

where the spectral phase-transfer kernel is

$$\delta\phi_i(\omega) \approx -\Re\left[\frac{H_E(\Omega, \omega)\tilde{E}_{s0}^*(\Omega)}{\tilde{E}_{i0}(\Omega)}\right] \tag{15}$$

Equation (14) is the central result of the linearized treatment. Once a reference operating point is chosen, the nonlinear DFWM process admits a well-defined spectral phase-response kernel for small perturbations.

To reduce this response to a one-dimensional performance curve analogous to the optical MTF, we probe the system with sinusoidal spectral-phase perturbations. Introducing a common mapped spectral coordinate uuu, we define the input phase modulation as

$$\delta\phi_s(u) \approx m_{in}\cos(\kappa u + \theta), \tag{16}$$

where $m_{in}$ is the input modulation depth, $\kappa$ is the spectral phase-modulation frequency, and $\theta$ is an arbitrary phase offset. In the locally mapped regime, the corresponding idler phase perturbation remains approximately sinusoidal at the same modulation frequency,

$$\delta\phi_i(u) \approx m_{out}(\kappa; \phi_{2,p}) \cos[\kappa u + \theta + \psi(\kappa; \phi_{2,p})]. \tag{17}$$

The minus sign reflects the complex-conjugated signal term in DFWM and therefore represents the trivial antisymmetric phase inversion rather than a loss of transfer fidelity.

We then define the complex SPMTF as

$$\mathcal{M}_\phi(\kappa; \phi_{2,p}) \approx \frac{m_{out}(\kappa; \phi_{2,p})}{m_{in}} \exp[i\psi(\kappa; \phi_{2,p})], \tag{18}$$

and its magnitude

$$M_\phi(\kappa; \phi_{2,p}) = |\mathcal{M}_\phi(\kappa; \phi_{2,p})| \approx \frac{m_{out}(\kappa; \phi_{2,p})}{m_{in}}. \tag{19}$$

We refer to $M_\phi(\kappa; \phi_{2,p})$ as the SPMTF. It provides a compact measure of how efficiently a phase modulation of spectral frequency $\kappa$ is transferred from the NIR signal to the UV idler for a given pump GDD. In analogy with the MTF in imaging, the SPMTF characterizes the effective resolution of the nonlinear phase-transfer platform in the spectral-phase domain: low-$\kappa$ modulations are expected to transfer faithfully, whereas high-$\kappa$ modulations should be progressively suppressed once the effective temporal overlap and mixing bandwidth become insufficient.

3. **Simulation**

We evaluate the SPMTF numerically by applying sinusoidal phase masks with different modulation frequencies to the input signal, propagating the fields through an idealized chirped-DFWM model, retrieving the idler phase, and projecting the result onto the same sinusoidal basis. Repeating this procedure for different pump GDD values yields a family of SPMTF curves, $M_\phi(\kappa; \phi_{2,p})$, which directly compare the phase-transfer bandwidth of different pump-chirp operating points.

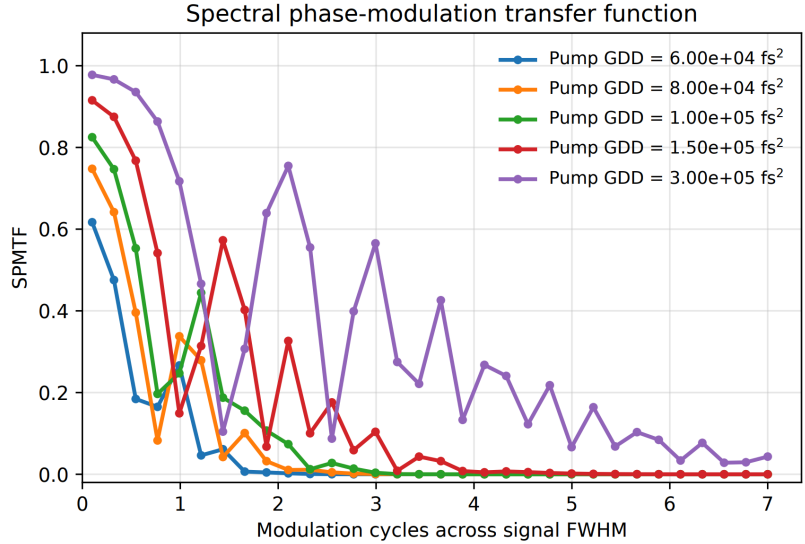

**Figure 1.** Calculated spectral phase-modulation transfer functions (SPMTFs) for several pump group-delay dispersion (GDD) values. The horizontal axis gives the number of phase-modulation cycles across the signal FWHM, and the vertical axis gives the transferred phase-modulation depth normalized to the input modulation depth. Larger pump GDD produces a broader SPMTF, indicating improved transfer of finer spectral-phase structure.

Figure 1 shows a clear trend: larger pump chirp broadens the response and preserves higher-frequency spectral-phase structure over a wider modulation range. Smaller GDD values show a rapid drop in $M_\phi$, indicating that only the slowly varying phase is transferred faithfully. Physically, increasing the pump chirp does not simply reshape the pump self-convolution in the frequency domain kernel; it also makes the mixing more localized in time and frequency. That improved selectivity increases the phase-transfer bandwidth, although only up to the point where the reduced peak intensity begins to penalize conversion efficiency. The SPMTF therefore makes the fidelity-efficiency trade-off visible in a single plot[5–8].

## 4. Conclusion

In summary, we recast chirped DFWM as a pump-conditioned spectral transfer kernel and show that, after linearization around a reference operating point, the signal-to-idler phase response can be described by a spectral phase-response kernel. This leads naturally to the SPMTF, obtained by probing the system with sinusoidal spectral-phase perturbations. The SPMTF provides a simple way to compare pump conditions and to quantify the phase-transfer bandwidth of a nonlinear frequency-conversion platform. Beyond the specific DFWM case considered here, the same framework should be useful whenever phase transfer is approximately local in a mapped spectral coordinate.